\documentclass[prd,twocolumn,superscriptaddress,showpacs,nofootinbib,amsmath,amssymb]{revtex4-1}
\usepackage{graphicx,bm}
\usepackage{color}
\definecolor{darkgreen}{RGB}{0,120,0}

\begin{document}
\title{Mapping dark matter in the gamma-ray sky with galaxy catalogs}
\author{Shin'ichiro Ando}
\affiliation{GRAPPA Institute, University of Amsterdam, 1098 XH
Amsterdam, The Netherlands}
\author{Aur{\'e}lien Benoit-L{\'e}vy}
\affiliation{Department of Physics and Astronomy, University College
London, London WC1E 6BT, United Kingdom}
\author{Eiichiro Komatsu}
\affiliation{Max-Planck-Institut f\"{u}r Astrophysik,
Karl-Schwarzschild Str. 1, 85741 Garching, Germany}
\affiliation{Kavli Institute for the Physics and Mathematics of the
Universe (Kavli IPMU, WPI), Todai Institutes for Advanced Study, The
University of Tokyo, Kashiwa, Chiba 277-8583, Japan}
\date{December 16, 2013; revised June 10, 2014}

\begin{abstract}
Cross-correlating gamma-ray maps with locations of galaxies in the
 low-redshift Universe {\it vastly} increases sensitivity to signatures
 of annihilation of dark matter particles. Low-redshift galaxies are ideal
 targets, as the largest contribution to anisotropy in the gamma-ray
 sky from annihilation comes from $z\lesssim 0.1$, where we expect
 minimal contributions from astrophysical sources such as
 blazars. Cross-correlating the five-year data of Fermi-LAT with the
 redshift catalog of the 2MASS survey can detect gamma rays from
 annihilation if dark matter has the canonical annihilation cross
 section and its mass is smaller than $\sim$100~GeV. 
\end{abstract}
\maketitle

\section{Introduction}

If dark matter particles annihilate and produce gamma rays, 
as predicted for a popular class of dark matter candidates,
we can detect them in {\it anisotropy} in the gamma-ray sky, since the
sites of annihilation trace the inhomogeneous matter distribution in the
Universe~\cite{APS}.
Gamma-ray anisotropy 
has been studied theoretically~\cite{APS, APS2, 
APS4, APS5, APS6, APS7, APS8, APS9, APS10, APS11, APS12, AP, APS13,
APS14, APS15, APS16, AK2013}
and recently
detected with the 22-month data of the Fermi Large Area
Telescope (LAT)~\cite{FermiAPS}.
This signal, however, can be explained entirely by active galaxies called
blazars~\cite{FermiInterpretation}. The current upper bounds on the rate
of dark matter annihilation obtained from gamma-ray anisotropy data are still
too weak to test the most interesting parameter regions, in which the dark
matter mass is on the order of 100~GeV and the annihilation cross
section is $\langle\sigma v \rangle = 3\times
10^{-26}$~cm$^{3}$~s$^{-1}$~\cite{AK2013,GomezVargas}
as implied from thermal production
mechanisms~\cite{Steigman:2012}. 

The current bounds are weak because the Fermi-LAT data at high energies
($\sim$10~GeV), which are sensitive to dark matter masses of
$\sim$100~GeV, are still totally photon-noise dominated. One way to
extract
signals more efficiently in such a low signal-to-noise regime is to {\it
cross-correlate} the noise-dominated data with some signal-dominated
data whose signals are spatially well-correlated with those in the
noise-dominated data. 

In this paper, we show that the {\it existing} catalog of locations of
galaxies in the low-redshift Universe ($z\lesssim 0.1$) measured by the
spectroscopic follow-up observations of the Two Micron All Sky Survey
(2MASS)~\cite{2MRS} provides an excellent template, which vastly
increases sensitivity to dark matter annihilation in cross correlation
with the Fermi-LAT data. 
We find that the expected sensitivity from the
five-year Fermi data is stringent enough to probe the most
interesting parameter regions of the annihilation cross section for
large ranges of dark matter masses, although the exact limits are still
subject to uncertainties on abundance of dark matter substructures.
We also show that the existing upper limits on the cross correlation of the
21-month Fermi data with galaxy catalogs~\cite{Xia2011} already yield 
much improved limits on the dark matter properties.
Cross correlation of the gamma-ray data with gravitational lensing data
can also be used to increase sensitivity to dark matter annihilation, as
shown by Ref.~\cite{CrossLensing}.

\section{Intensity of the diffuse gamma-ray background}

The mean intensity of the diffuse gamma-ray background due to dark matter
annihilation (the number of photons received per unit area, time, solid
angle, and energy range) is given by
\begin{equation}
I_{\rm dm}(E) = \int d\chi W_{\rm
dm}([1+z]E, \chi)\langle \delta^2   (z) \rangle,
\label{eq:Idm}
\end{equation}
 where $\chi$ is the
comoving distance out to a given redshift $z$. A window function,
\begin{equation}
W_{\rm dm}(E,z) = \frac{\langle \sigma v\rangle}{8\pi}
 \left(\frac{\Omega_{\rm dm}\rho_c}{m_{\rm dm}}\right)^2 (1+z)^3
 \frac{dN_{\gamma,{\rm ann}}}{dE} e^{-\tau}, 
\end{equation}
 contains all the particle-physics information such as
the annihilation cross section, $\langle \sigma v \rangle$, the dark
matter mass, $m_{\rm dm}$, the energy spectrum of emitted gamma rays per
annihilation, $dN_{\gamma,{\rm ann}}/dE$, and the optical depth of
absorption during propagation in the intergalactic space, $\tau (E,z)$
(e.g.,~\cite{EBL}). We use $\Omega_{\rm dm} = 0.23$ for the density
parameter of dark matter, and $\rho_c$ is the present critical
density of the Universe. The variance of the matter density fluctuation,
$\langle \delta^2 \rangle$, is given by  
\begin{eqnarray}
 \langle \delta^2\rangle &=& \left(\frac{1}{\Omega_m\rho_c}\right)^2
  \int dM \frac{dn(M,z)}{dM} [1+b_{\rm sh}(M)]
  \nonumber\\&&{}\times
  \int dV \rho_{\rm host}^2(r|M),
\label{eq:delta2}
\end{eqnarray}
where $dn/dM$ is the comoving number density of dark matter halos per
unit mass range, $\rho_{\rm host}(r|M)$ is the density profile of dark
matter halos of mass $M$, and $b_{\rm sh} (M)$ is the so-called ``boost
factor'' due to the presence of subhalos inside parent dark matter
halos. See Ref.~\cite{AK2013} for how to evaluate
Eq.~(\ref{eq:delta2}) as well as for the cosmological parameters used in
the calculation. The boost factor, $b_{\rm sh}$, depends on the minimum
mass of possible subhalos as well as on the host halo mass. For the
minimum subhalo mass, we use the standard value for the cold dark matter
particles, $10^{-6} M_\odot$.

Let us specify some important details of the model. As the rate of
annihilation depends on local density squared, the results are sensitive
to how clumpy dark matter halos are. There are two important quantities
related to clumpiness. One is the so-called ``concentration parameter''
of the density profile of halos, and we use the model developed in
Ref.~\cite{bullock} for $M<2.5\times 10^{14} M_\odot$ and that in
Ref.~\cite{duffy} otherwise.  This model yields results
similar to the latest work~\cite{ludlow}. We find that the most
of the contributions to the anisotropy come from subhalos inside the
large-mass halos ($M\gtrsim 10^{10} M_\odot$) at low redshifts
($z\lesssim 0.1$)~\cite{AK2013}. The concentration parameters of such
large-mass halos have been well characterized. Another important
quantity is the boost factor due to subhalos, and we use the model
developed by Gao et al.~\cite{gao}. Their power-law scaling with mass,
$b_{\rm sh}\propto M^{0.39}$, is recently
challenged by S\'anchez-Conde and Prada~\cite{prada2}, who claim to find
significantly weaker dependence of $b_{\rm sh}$ on $M$. This greatly
reduces the amplitude of anisotropies as well as the mean
intensity. 
While we continue to adopt the model of
Ref.~\cite{gao} as the main model in this paper, our conclusion
changes if the model of Ref.~\cite{prada2} turns out to be
correct. This is the largest uncertainty in our model,
and is common to {\it all} the extragalactic constraints discussed in
the literature.

\begin{figure}[t]
 \begin{center}
  \includegraphics[width=8.5cm]{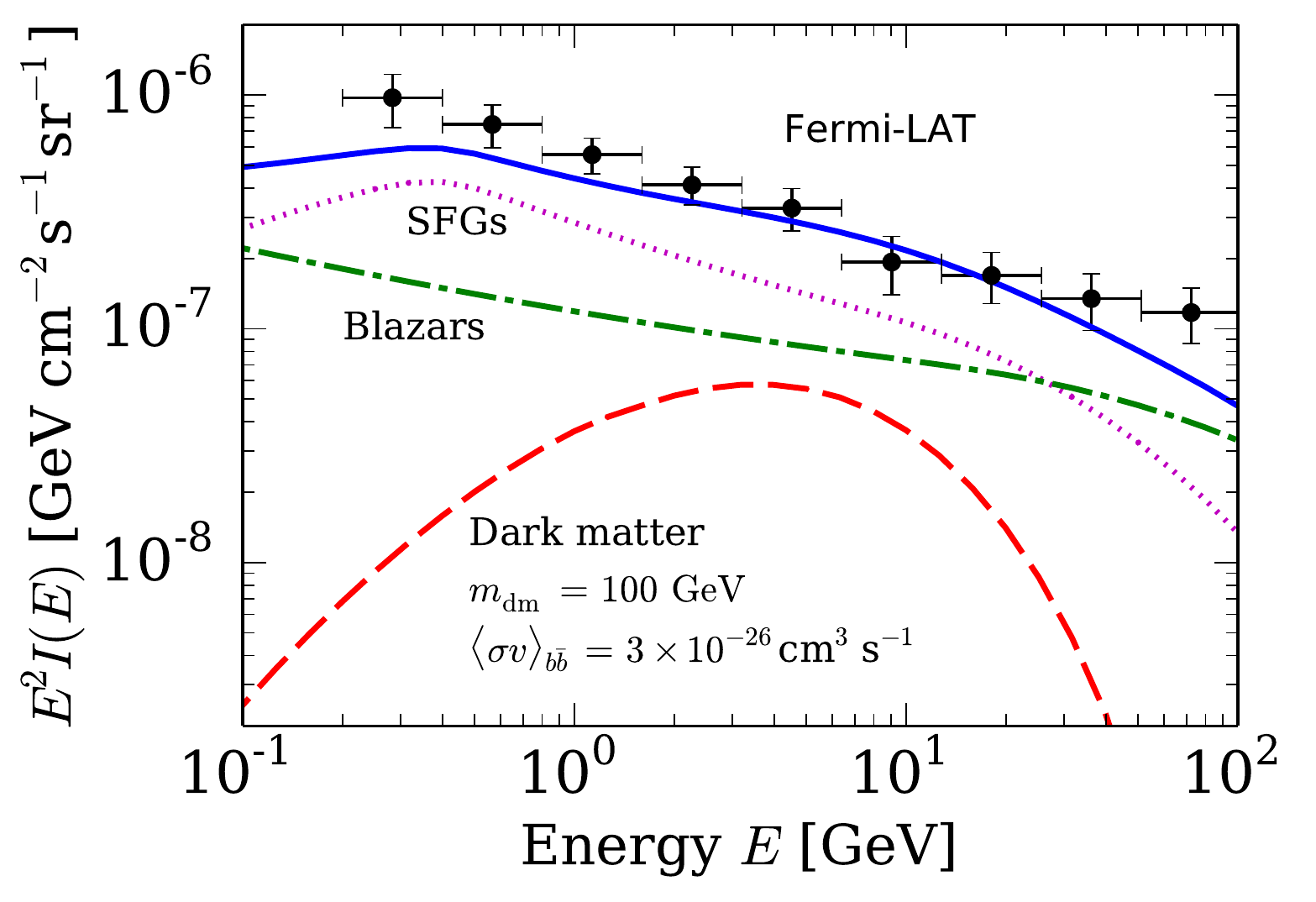}
  \caption{Predicted mean intensity spectra of diffuse gamma-ray background. The
  dashed, dot-dashed, and dotted lines show the contributions from dark
  matter annihilation (parameters adopted are also shown), blazars, and
  star-forming galaxies (labeled as SFGs),
  respectively. The solid line shows the sum, while the points with
  error bars show the Fermi-LAT data~\cite{FermiDiffuse}.}
  \label{fig:Idm}
 \end{center}
\end{figure}

In Fig.~\ref{fig:Idm}, we show $I_{\rm dm}(E)$ from annihilation of
100-GeV dark matter purely into $b\bar b$ with $\langle \sigma v\rangle
=  3\times 10^{-26}~\mathrm{cm^3~s^{-1}}$, as well as from
two astrophysical sources: blazars and star-forming galaxies.
For both populations, we treat spectrally hard and soft sub-populations,
separately: BL Lacs ($E^{-2.1}$) and flat-spectrum radio quasars
($E^{-2.4}$) for the blazars; starbursts ($E^{-2.2}$) and normal spirals
($E^{-2.7}$) for the star-forming galaxies.
The mean intensity of these sources is computed in a
similar manner to dark matter, using Eq.~(\ref{eq:Idm}) but by replacing
$W_{\rm dm}$ with a window function of each population and $\langle
\delta^2\rangle$ with 1 (as they trace density).
The window function $W_X$, where $X$ represents either star-forming
galaxies or blazars, is given by
\begin{equation}
 W_X([1+z]E,z) = \chi^2 \int_0^{\mathcal L_{\rm lim}}
  d \mathcal L \Phi_X (\mathcal L, z) \mathcal F_X(\mathcal L,z),
  \label{eq:W_X}
\end{equation}
where $\mathcal L$ is the differential luminosity (i.e., the number of
gamma-ray photons emitted per unit time, per unit energy range) at energy
$(1+z)E$, and $\mathcal F_X(\mathcal L, z) = (1+z)^2 \mathcal L/(4\pi
d_L^2)$ is the differential number flux at energy $E$ from a source $X$
at $z$.
The upper limit of the integration $\mathcal L_{\rm lim}$ corresponds to
the flux sensitivity of Fermi, $F_{\rm lim}$, integrated above 100~MeV,
and we adopt $F_{\rm lim} = 4\times 10^{-9}$~cm$^{-2}$~s$^{-1}$
($3\times 10^{-8}$~cm$^{-2}$~s$^{-1}$) for hard (soft) sources.
For the luminosity function, $\Phi_X$, for blazars, we adopt the
luminosity-dependent density evolution model separately for BL
Lacs~\cite{LDDE1} and flat-spectrum radio quasars~\cite{LDDE2}, which
both roughly behave as a broken power law in luminosity.
For the lumonosity function of star-forming galaxies, we adopt the
infrared luminosity function~\cite{IR}, which behaves as a power law
with a cutoff luminosity, again separately for spiral and starburst
galaxies. Finally such an infrared luminosity function is converted to
the gamma-ray luminosity function by using the correlation between
infrared and gamma-ray lumonosity calibrated with Fermi: $L_\gamma
\propto L_{\rm IR}^{1.17}$~\cite{FermiGal}.

Figure~\ref{fig:Idm} shows that the annihilation signal is below the
current measurements as well as the predicted astrophysical
contributions.
This situation, however, changes completely when we consider the cross
correlation of anisotropies.

\section{Cross correlation with 2MASS galaxy catalog}
\label{sec:cross correlation}

\begin{figure}[t]
 \begin{center}
  \includegraphics[width=8.5cm]{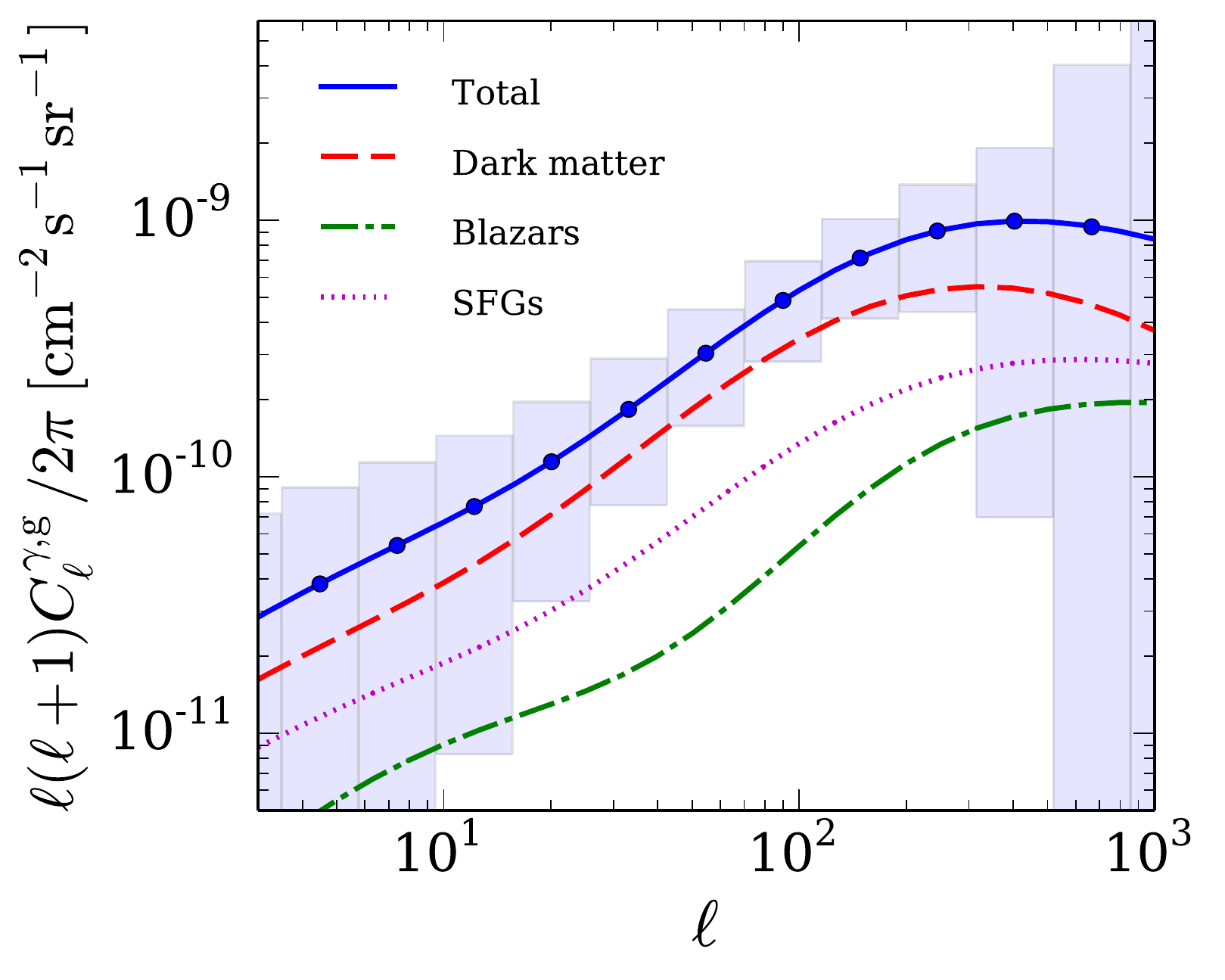}
  \caption{Predicted angular cross-power spectra of gamma-ray emission
  in 5--10~GeV and the distribution of galaxies measured by the 2MASS
  Redshift Survey. The dashed, dot-dashed, and dotted lines show the
  contributions from dark matter annihilation, blazars, and star-forming
  galaxies, respectively. The solid line shows the sum, while the points
  with the boxes show the errors expected after five-year observations of
  Fermi-LAT. The particle physics model is the same as in
  Fig.~\ref{fig:Idm}.}
  \label{fig:Cl_2MRS}
 \end{center}
\end{figure}

We consider the {\it cross-correlation power spectrum},
$C^{\rm dm,g}_\ell$, between the fluctuations in the gamma-ray
intensity $\delta I_{\rm dm}$, and the galaxy density contrast
$\delta_{\rm g}$. It is defined by
\begin{equation}
\langle\delta I_{\rm dm}(\hat{\bm
n})\delta_{\rm g}(\hat{\bm n}+{\bm \theta})\rangle= 
\sum_\ell \frac{2\ell+1}{4\pi}C^{\rm dm,g}_\ell P_\ell(\cos\theta),
\end{equation}
where $P_\ell(\cos\theta)$ is the Legendre polynomials.
Each multipole roughly corresponds to an angular size of $\theta \approx
\pi/\ell$. We compute $C^{\rm dm,g}_\ell$ as 
\begin{equation}
 C_{\ell}^{\rm dm, g} = \int
  \frac{d\chi}{\chi^2} W_{\rm dm}(\chi) W_{\rm g}(\chi) P_{\delta^2,
  {\rm g}}\left(k=\frac{\ell}{\chi}, \chi\right),
  \label{eq:angular cross-power}
\end{equation}
where $W_{\rm g}$ is the galaxy window function, normalized to unity after
integration over $\chi$. The angular cross-power spectrum is determined
by the three-dimensional cross-power spectrum of $\delta^2$ and
galaxies, $P_{\delta^2, {\rm g}}(k)$. We model this power spectrum as
$P_{\delta^2, {\rm g}}(k) = b_{\rm g} P_{\delta^2, \delta}(k)$, where
$b_{\rm g}$ is the so-called galaxy bias factor. We use  $b_{\rm g}=
1.4$ for galaxies in the 2MASS catalog~\cite{Davis:2011}.

To compute $P_{\delta^2, \delta}(k)$, we extend the formalism given in
Ref.~\cite{AK2013} to the cross correlation and obtain
$P_{\delta^2,\delta} = P_{\delta^2, \delta}^{1h} +
P_{\delta^2,\delta}^{2h}$, where 
\begin{eqnarray}
 P_{\delta^2, \delta}^{1h} &=& \left(\frac{1}{\Omega_m\rho_c}\right)^3
  \int dM \frac{dn}{dM}\tilde u(k|M) \tilde v(k|M) M
  \nonumber\\&&{}\times
  [1+b_{\rm sh}(M)]  \int  dV \rho_{\rm host}^2(r|M),\\
 P_{\delta^2, \delta}^{2h} &=& \left(\frac{1}{\Omega_m\rho_c}\right)^2
  \left\{\int dM \frac{dn}{dM} \tilde u(k|M) b_1(M,z)
   \right.\nonumber\\&&{}\left.
   \times[1+b_{\rm sh}(M)] \int dV \rho_{\rm host}^2(r|M)\right\}
   \nonumber\\&&{}\times
   \left[\int dM \frac{dn}{dM} M \tilde v(k|M) b_1(M,z)\right]
   P_{\rm lin}(k,z),
\end{eqnarray}
where $P_{\rm lin}(k,z)$ is the linear matter power spectrum, $b_1(M,z)$
is the linear halo bias, and $\tilde u(k|M)$ and $\tilde v(k|M)$ are the
Fourier transform of gamma-ray emissivity and density profiles,
respectively, which are both normalized to unity after integration
over volume.

For the cross correlation of the astrophysical sources with 2MASS
galaxies, we use Eq.~(\ref{eq:angular cross-power}) with a proper
replacement of $W_{\rm dm}$ with the astrophysical window function
[Eq.~(\ref{eq:W_X})].
We also replace the power spectrum $P_{\delta^2, {\rm g}}$
with $P_{X, {\rm g}}$, and we approximate it as $P_{X, {\rm g}} \approx
b_X b_{\rm g} P_\delta$, where $P_\delta$ is the matter power spectrum.
For both blazars and star-forming galaxies, we assume $b_X = 1.4$ for
their bias parameters.

The angular power spectrum defined by Eq.~(\ref{eq:angular cross-power})
has units of intensity times solid angle, and it is proportional to 
$\langle \sigma v \rangle$. In Fig.~\ref{fig:Cl_2MRS}, we show the
predicted $C_\ell^{{\rm dm}, {\rm g}}$ with the 2MASS Redshift
Survey~\cite{2MRS}, assuming $\langle \sigma v \rangle = 3\times
10^{-26}$~cm$^{3}$~s$^{-1}$ in the energy range of 5--10~GeV, for 100-GeV
dark matter annihilating into $b\bar b$. We also show the
predicted cross spectra with the 2MASS Redshift Survey for
blazars and star-forming galaxies,
respectively. Remarkably, we find that the dark matter-galaxy
correlation dominates over the other astrophysical contributions.
This is because the low-redshift ($z\lesssim 0.1$) 2MASS galaxies are less
correlated with the astrophysical gamma-ray sources than with dark
matter annihilation. 
The galactic emission due to cosmic ray
interactions is much more concentrated at the halo center than dark 
matter annihilation; thus, while the former is easier to be identified 
with nearby individual sources, the latter yields the larger luminosity 
density in a local volume.\footnote{
For example, several star-forming galaxies in the local volume such as
M31 were detected with Fermi, and they are all consistent with
cosmic-ray origin~\cite{FermiGal}.
This, however, does {\it not} exclude the possibility that the
entire local volume contains more gamma rays due to dark matter
annihilation than to cosmic rays.
This is because, even if a galaxy identified with Fermi emits the same
amount of photons from dark matter annihilation, they give much
smaller surface brightness, and hence would not be detected, as dark
matter is distributed out to a virial radius.}
It is therefore important to use a local galaxy
catalog such as 2MASS to reduce contamination from blazars and
star-forming galaxies.

\begin{figure}[t]
 \begin{center}
  \includegraphics[width=8.5cm]{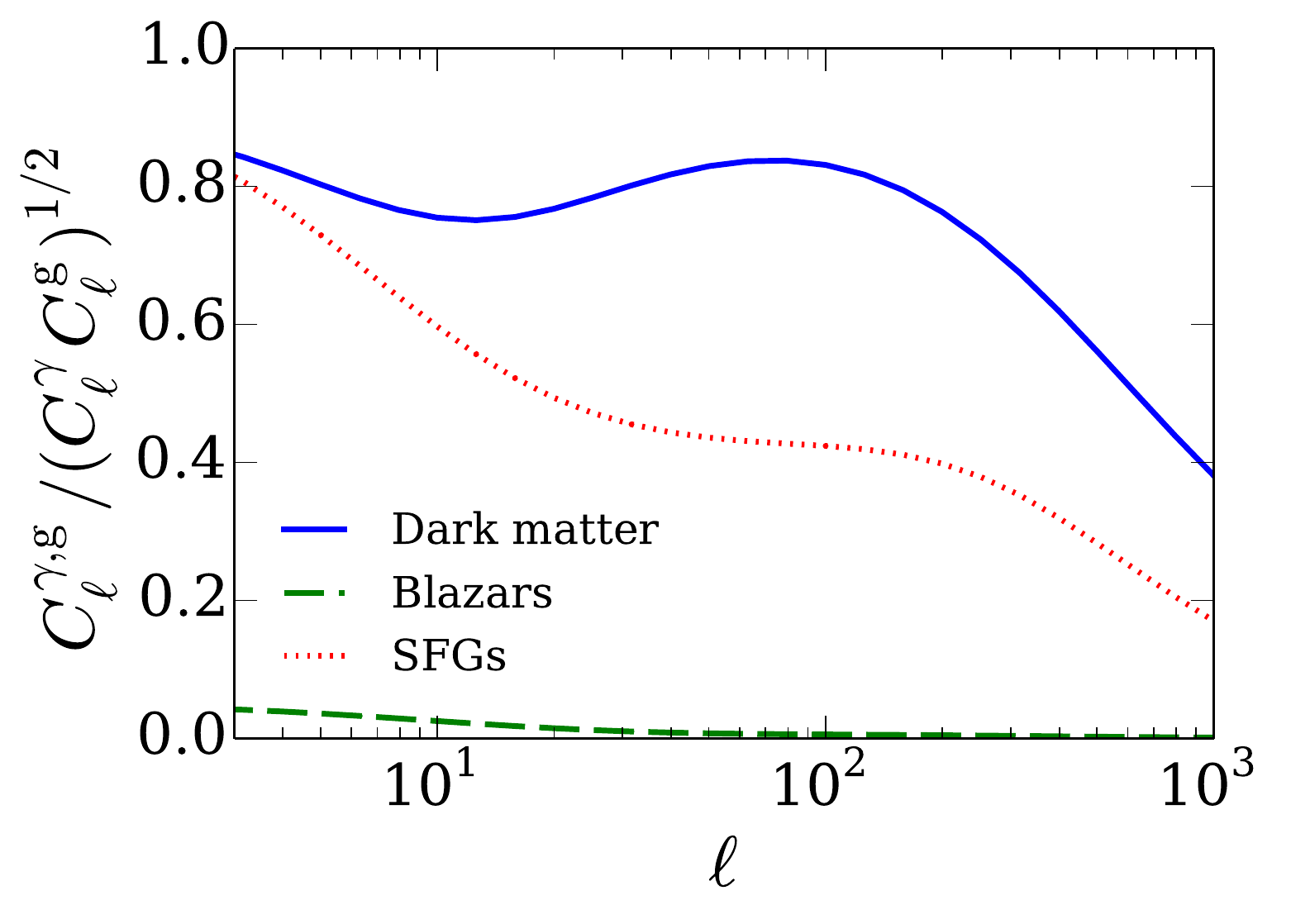}
  \caption{Predicted cross-correlation coefficients,
  $C_\ell^{\gamma,{\rm g}}/\sqrt{C_\ell^\gamma C_\ell^{\rm g}}$, between gamma rays
  from dark matter (solid), blazars (dashed), or star-forming galaxies
  (dotted), and the 2MASS Redshift Survey galaxies.}
  \label{fig:cross_coeff}
 \end{center}
\end{figure}

In Fig.~\ref{fig:cross_coeff}, we show the cross-correlation
coefficients, $C_\ell^{\gamma,{\rm g}}/\sqrt{C_\ell^\gamma C_\ell^{\rm
g}}$, between gamma rays from dark matter, blazars, or star-forming
galaxies, and the 2MASS galaxies. We find that the gamma rays from dark
matter annihilation
and the 2MASS galaxies are spatially well-correlated, with the
cross-correlation coefficients of about $0.8$ over a wide range of
multipoles up to $\ell\approx 200$. On the other hand, blazars are
poorly correlated, having negligible cross-correlation
coefficients. This is because, as far as we know now, there are
not many blazars in the local Universe (with the assumed lowest
gamma-ray luminosity of $10^{42}$~erg~s$^{-1}$ above
100~MeV). Star-forming
galaxies are also less correlated, having the cross-correlation
coefficients of $\lesssim 0.5$ at $\ell\gtrsim 10$.

Another remarkable finding from Fig.~\ref{fig:Cl_2MRS} is that the
existing data, i.e., the five-year data of Fermi-LAT and the galaxies in
the 2MASS Redshift Survey, have sufficient sensitivity to detect
gamma rays from dark matter annihilation with the canonical annihilation
cross section and $m_{\rm dm}=100$~GeV. To compute the predicted error
bars, we use 
\begin{eqnarray}
 \delta C_\ell^{\gamma, {\rm g}} &=& \sqrt{\frac{1}{(2\ell +1) f_{\rm sky}}}
  \Biggl[(C_\ell^{\gamma, {\rm g}})^2 
   \nonumber\\&&{}+
  \left(C_\ell^\gamma + \frac{C_N^\gamma}{W_\ell^2}\right)
  \left(C_\ell^{\rm g} + C_N^{\rm g}\right)\Biggr]^{1/2},
  \label{eq:error}
\end{eqnarray}
where $f_{\rm sky}=0.7$ is a fraction of the sky used for the analysis.
For the auto-correlation power spectrum, $C_\ell^\gamma$, we use the
measured values reported in Ref.~\cite{FermiAPS}, and the photon noise
is estimated as $C_N^\gamma = I_{\rm obs} / \mathcal E$, where $I_{\rm
obs}$ is the
observed mean intensity reported in Ref.~\cite{FermiDiffuse} and $\mathcal E
= 1.5 \times 10^{11}$~cm$^2$~s is the exposure for the
five-year Fermi-LAT operation (almost independent of energy). As for the
window function of the Fermi-LAT angular response, $W_\ell$, we use a
functional form that approximates the results reported in
Ref.~\cite{FermiAPS}. It is straightforward to calculate the angular
power spectrum of the galaxies, $C_\ell^{\rm g}$, from the redshift
distribution of galaxies in the 2MASS Redshift Survey (see,
e.g.,~\cite{AP}). Finally, $C_N^{\rm g}$ is the shot noise of galaxies
given by $C_N^{\rm g} = 4\pi f_{\rm 2MASS} / N_{\rm g}$, where $N_{\rm g}
= 43500$ is the number of 2MASS galaxies with measured redshifts over
$f_{\rm 2MASS} = 0.91$ of the sky. 

A simple error propagation gives the expected uncertainties on theoretical
parameters, $\{\vartheta_a\}$, given the errors on the power spectrum. We
first compute the Fisher matrix
\begin{equation}
F_{ab} = \sum_\ell \sum_i
 \frac{(\partial C_{\ell,i}^{\gamma, {\rm g}} / \partial
 \vartheta_a)(\partial C_{\ell,i}^{\gamma, {\rm g}} / \partial \vartheta_b)}
 {(\delta C_{\ell,i}^{\gamma, {\rm g}})^2},
\end{equation}
where $\vartheta_1 = \langle \sigma v\rangle$ and $\vartheta_{2}$ and
$\vartheta_3$ are the amplitudes of the cross spectra for the star-forming
galaxies and blazars, respectively, the subscript $i$ represents four
energy bands (1--2, 2--5, 5--10, and 10--50~GeV), and $\delta
C_{\ell,i}^{\gamma, {\rm g}}$ in the denominator is evaluated using
Eq.~(\ref{eq:error}) at $\langle \sigma v \rangle = 3\times
10^{-26}$~cm$^{3}$~s$^{-1}$.\footnote{While the error, $\delta C_{\ell,i}^{\gamma, {\rm g}}$, also depends on $\langle
\sigma v \rangle$, the dependence is weak because it is dominated by the
photon shot noise. We thus fix $\langle \sigma v \rangle$ to be the canonical value in $\delta C_{\ell,i}^{\gamma, {\rm g}}$ when computing the Fisher matrix.}
The covariance between different energy
ranges is negligible, as the auto-power spectrum of the gamma-ray data 
is dominated by shot noise of the photons.
The 95\%~CL upper bound on $\langle \sigma v \rangle$ is then obtained
with $\langle \sigma v \rangle < 1.64 \sqrt{(F^{-1})_{11}}$.

\begin{figure}[t]
 \begin{center}
  \includegraphics[width=8.5cm]{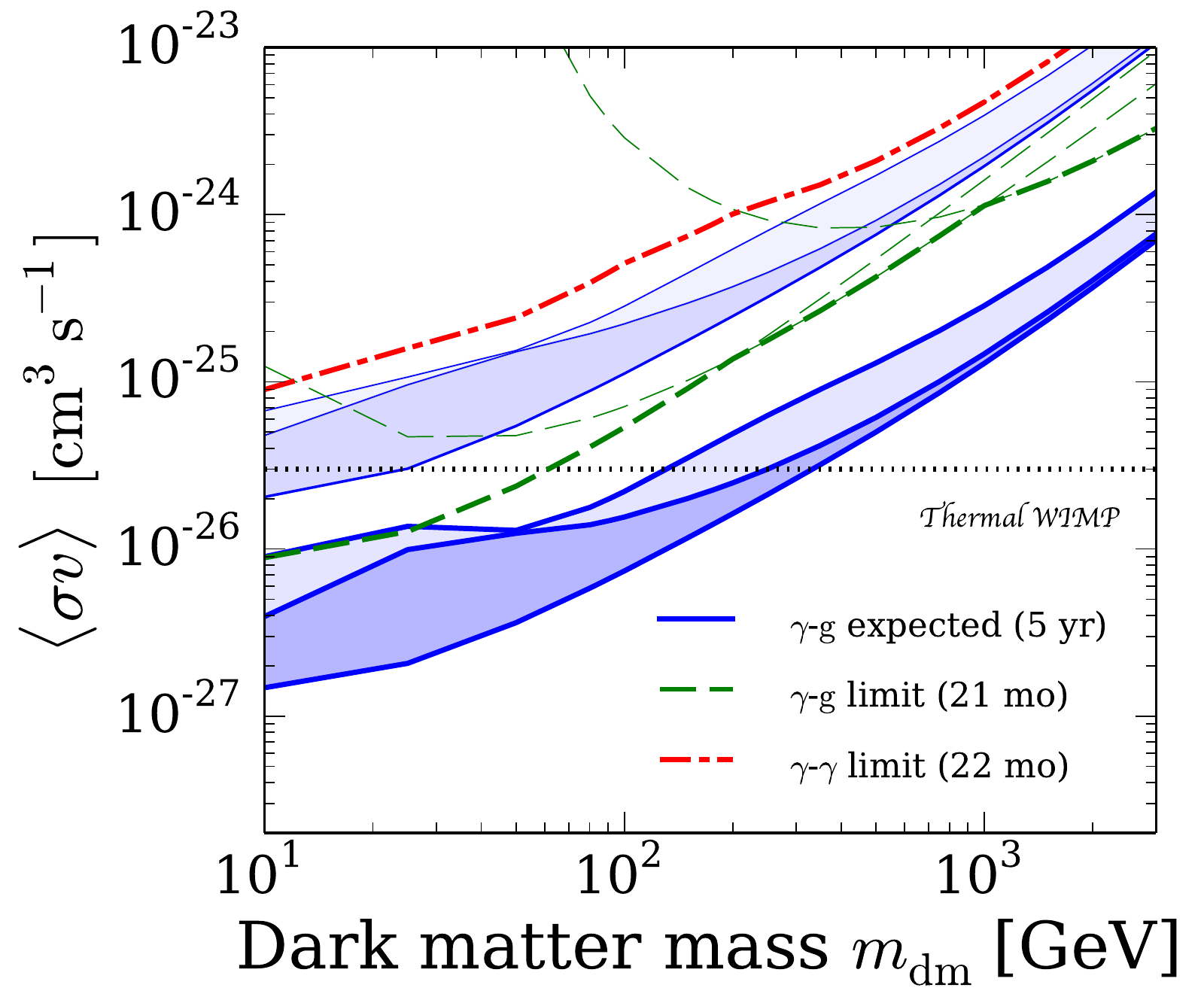}
  \caption{Existing and expected 95\%~CL upper bounds on the cross
  section of annihilation into $b\bar b$. The dot-dashed line shows the
  bounds from the auto power spectrum of the 22-month of
  Fermi-LAT~\cite{AK2013}, while the dashed lines show the bounds from
  the cross-power spectra between the 21-month of Fermi-LAT ($E>1$, 3,
  and 30~GeV from the left to right) and the 2MASS galaxies (with
  photometric redshifts) that we derive using the measurement reported
  in Ref.~\cite{Xia2011}.
  The solid curves show the expected bounds from
  the five-year Fermi-LAT data cross-correlated with galaxies in the
  2MASS Redshift Survey, based on the assumption that all the
  astrophysical contributions (lower), only blazars (middle), or none of
  them (upper) are understood. The lower, thick solid curves are derived
  using the boost factor model of Ref.~\cite{gao}, whereas the upper,
  thin solid curves are based on Ref.~\cite{prada2}.
  }
  \label{fig:sigmav_limit_2MRS}
 \end{center}
\end{figure}

The lowest thick solid curve in
Fig.~\ref{fig:sigmav_limit_2MRS} shows the most 
optimistic constraint obtained by assuming that we know the
amplitudes of contributions from star-forming galaxies and blazars,
i.e., $\langle \sigma v \rangle < 1.64/\sqrt{F_{11}}$. The second to the
lowest curve shows the realistic bound obtained by varying the
amplitudes of contributions from dark matter and star-forming galaxies
(i.e., the Fisher matrix is a $2 \times 2$ matrix for $\vartheta_1$ and
$\vartheta_2$).
Finally, the upper thick solid curve shows the conservative bound
obtained by varying all $\vartheta_a$'s;
i.e., this bound is obtained by using no prior knowledge on the
amplitudes.
This is conservative because we know that the blazar contribution is
tightly constrained to be negligible from its luminosity
function~\cite{LDDE1, LDDE2} and the angular power
spectrum~\cite{FermiInterpretation}, and thus the error associated
to this component is much smaller.


The existing bounds on $\langle\sigma v\rangle$ from the auto
power spectra of the 22-month Fermi-LAT~\cite{AK2013} data are shown in
the dot-dashed line.
The five-year cross spectra can improve the constraints by more than one
order of magnitude, testing the most interesting parameter space of the
dark matter masses and cross sections.

One partial reason for the improvement over the
auto-power spectrum is 
that the analysis of the auto power spectrum is limited to $\ell\ge 155$
due to potential contamination by the diffuse Galactic emission
\cite{FermiAPS}, whereas we can use the entire multipoles for the
cross-power spectrum, as the Galactic emission is not correlated with the
locations of 2MASS galaxies.

This conclusion is subject to the theoretical uncertainty
regarding the boost factor, $b_{\rm sh}(M)$.
The model of Ref.~\cite{gao} relies on an extrapolation of a power-law
relation between the concentration parameter and the halo mass.
Reference~\cite{prada2} argues that such a power-law extrapolation to
very small masses is unphysical, and the relation must necessarily
flatten toward lower masses, resulting in substantially fainter subhalos
in parent halos with $M\gtrsim 10^{10} M_\odot$.
If we adopt the latter model~\cite{prada2}, the amplitude of the cross
correlation shown in Fig.~\ref{fig:Cl_2MRS} gets reduced, while there is
little change in its shape.
The thin solid curves in Fig.~\ref{fig:sigmav_limit_2MRS} shows the 
expected five-year constraints using the same model.
In this case the constraints weaken by an order of magnitude.
This represents the current theoretical uncertainty in the prediction.

\section{21-month upper limits from cross-correlation study}
\label{sec:data}

What do the current data tell us? Xia et al.~\cite{Xia2011} have
measured the cross-correlation function in configuration space,
$C(\theta)$, from the 21-month Fermi-LAT data and 770000 2MASS galaxies
with photometric (rather than spectroscopic) redshifts over the angular
range of one to ten degrees. They see no evidence for cross correlation.
Thus, we use our model to place upper bounds on $\langle\sigma v\rangle$
from their measurements, by fixing components of the star-forming
galaxies and blazars to their baseline models. We compute the $\chi^2$
statistics: 
\begin{equation}
 \chi^2 = \sum_{ij} [C^{\gamma, {\rm g}} (\theta_i) - C_i](\delta C^2)_{ij}^{-1}
  [C^{\gamma, {\rm g}}(\theta_j)  - C_j],
\end{equation}
where $C^{\gamma, {\rm g}} (\theta)$ is the theoretical value of the
cross-correlation function corresponding to $C_\ell^{\gamma, {\rm g}}$,
$C_i$ is the measured cross correlation in the
$i$-th angular bin, and $(\delta C^{2})_{ij}^{-1}$ is the inverse
covariance matrix computed from the jackknife analysis~\cite{Xia2011}.
The dashed lines in Fig.~\ref{fig:sigmav_limit_2MRS} show the 95\%~CL
upper bounds on $\langle\sigma v\rangle$ from the measurements in $E>1$,
3, and 30~GeV. The existing data indeed provide a significant
improvement over the auto power spectrum. This analysis thus provides a
proof of concept.

\section{Discussion and conclusions}

{\it Other possible astrophysical sources.---}%
There are other astrophysical source classes that trace dark matter and
contribute to the gamma-ray background.
Misaligned active galaxies are one such example, and it has been shown
that they could give larger contribution to the mean intensity than
blazars although with much larger uncertainties (see, e.g.,
Ref.~\cite{MisalignedAGN}).
Our Fisher-matrix approach adopted in Sec.~\ref{sec:cross correlation},
however, relies on no prior information on the amplitudes of the
cross-power spectrum for both the blazars and star-forming galaxies,
with which we still obtain quite stringent upper limits on dark matter
annihilation cross section.
This is because the energy dependence on the cross-power spectrum is
properly taken into account, where a characteristic spectrum of dark
matter component (Fig.~\ref{fig:Idm}) plays an essential role.
Therefore, unless they feature an energy spectrum very different from a
power law, contributions from any other astrophysical sources will be
degenerate with either blazars or star-forming galaxies, and our
conclusions will be unchanged.

{\it Uncertainty on galaxy bias.---}%
The bias parameter of the galaxy distribution is well constrained in the
linear regime, but to a less extent in the nonlinear regime.
To this end, the approximation we adopted, $\delta_{\rm g} = b_{\rm g}
\delta$ with a constant linear bias $b_{\rm g}$, is appropriate for the
two-halo term, which dominates up to $\ell \sim 50$.
The one-halo term dominates at higher multipoles, where the galaxy bias
is likely nonlinear.
The nonlinear bias tends to increase the small-scale power spectrum,
which is currently difficult to predict due to uncertainties in the
physics of galaxy formation.
We thus assume the linear bias on small scales, but our prediction for
the small-scale power may be an underestimate by a few tens of percent.
However, this uncertainty on the bias for 2MASS galaxies unlikely
affects the conclusions, because it is common for all the gamma-ray
sources and we obtain significant statistics already from the linear
regime alone.

{\it Correlation with gravitational lensing.---}%
Instead of using galaxy positions, the cross-correlation analysis can be
performed with a map of projected mass-density fields constructed from
measurements of the gravitational lensing effect toward many background
galaxies~\cite{CrossLensing}. Such an analysis should also be pursued in
order to further increase sensitivity as well as to improve robustness
of the results against systematic errors. 

{\it Summary.---}%
We showed that taking cross correlations between the
gamma-ray background and the 2MASS galaxy catalog provides very strong
probe of dark matter annihilation.
If the mass of the dark matter particles is on the order of 100 GeV,
an upper limit on the annihilation cross section one can obtain by using
five-year Fermi data will exclude its canonical value inferred from the
thermal freeze-out scenario even after taking uncertainties on
astrophysical sources into account.
We note, however, that this depends on the amount of substructures
present in host dark matter halos.



\acknowledgments

We acknowledge the GRAPPA workshop on ``Anisotropic Universe'' for
providing the opportunity to initiate this study.
We would like to thank J.-Q.~Xia and A.~Cuoco for providing us with the
measurements of the cross-correlation function and its covariance matrix.
This work was supported by NWO through Vidi Grant (S.A.) and the
Leverhulme Trust and STFC (A.B.-L.).

\end{document}